\documentclass[aps,prb, twocolumn,superscriptaddress, noeprint]{revtex4-2}

\usepackage{comment}

\usepackage{hyperref}
\usepackage[utf8]{inputenc}
\usepackage[T1]{fontenc}    %
\usepackage[english]{babel}
\usepackage[pdftex]{graphicx}
\usepackage{amsmath,amssymb,amsthm,bbm, mathtools} %

\usepackage{slashed}
\usepackage{tikz}   %
\usepackage[left=2.5cm, right=2.5cm, top=2.5cm, bottom=2.5cm]{geometry}
\usepackage{multirow} %

\usepackage{placeins}%
                  
\numberwithin{equation}{section}

\renewcommand{\vec}{\boldsymbol}
\newcommand{\shortminus}{\text{-}}
\newcommand{\unit}[1]{\,\mathrm{#1}}

\begin{document}

\title{Magnetic ordering tendencies in hexagonal boron nitride-bilayer graphene moiré structures}
\author{Maria Spethmann} 
\affiliation{Institute for Theoretical Solid State Physics,
RWTH Aachen University, and JARA Fundamentals
of Future Information Technology, 52056 Aachen, Germany}
\author{Carsten Honerkamp}
\affiliation{Institute for Theoretical Solid State Physics,
RWTH Aachen University, and JARA Fundamentals
of Future Information Technology, 52056 Aachen, Germany}
\author{Dante M. Kennes}
\affiliation{Institute for Theory of Statistical Physics,
RWTH Aachen University, and JARA Fundamentals
of Future Information Technology, 52056 Aachen, Germany}
\affiliation{Max Planck Institute for the Structure and Dynamics of Matter, Center for Free Electron Laser Science, Luruper Chaussee 149, 22761 Hamburg, Germany}
\date{May 29, 2021}

\begin{abstract}
When hexagonal boron nitride (hBN) and graphene are aligned at zero or small twist angle, a moiré structure is formed due to the small lattice constant mismatch between the two structures.
In this work, we analyze magnetic ordering tendencies, driven by onsite Coulomb interactions, of encapsulated bilayer graphene (BG) forming a moiré structure with one (hBN-BG) or both hBN layers (hBN-BG-hBN), using the random phase approximation. The calculations are performed in a fully atomistic Hubbard model that takes into account all $\pi$-electrons of the carbon atoms in one moiré unit cell. 
We analyze the charge neutral case and find that the dominant magnetic ordering instability is uniformly antiferromagnetic. Furthermore, at low temperatures, the critical Hubbard interaction $U_c$ required to induce magnetic order is slightly larger in those systems where the moiré structure has caused a band gap opening in the non-interacting picture, although the difference is less than 6\%. Mean-field calculations are employed to estimate how such an interaction-induced magnetic order may change the observable single-particle gap sizes.
\end{abstract}
\maketitle

\section{Introduction}

hBN is frequently used as a substrate for graphene devices, because it offers a flat and clean surface and enables a high charge carrier mobility \cite{Yankowitz2019}. When single layer graphene (SG) and hBN are stacked at no or small twist angle to each other, they form a moiré structure due to a small lattice constant mismatch. hBN has a lattice constant of around $a_{hBN}=0.2504\unit{nm}$ \cite{liu2003} whereas SG has a lattice constant of $a=0.246\unit{nm}$ \cite{McCann2013}. The formed moiré structure, with lattice constants up to $14\unit{nm}$ \cite{Yankowitz2012}, changes the electronic properties of the material and has been analyzed in several theoretical \cite{slotman2015,jung2015,song2013,Moon2014} and experimental \cite{woods2014,ponomarenko2013,hunt2013} papers. 
For example, a band gap opens at the central Dirac points of SG \cite{hunt2013}. %
The size of the central band gap was analyzed in calculations, which stressed the importance of structural relaxation of the carbon atoms within the moiré structure, but also included many-body perturbation theory and electron-electron interactions \cite{Bokdam2014,jung2015}.
Further, secondary, gapped Dirac points have been reported to form at the mini Brillouin zone edges \cite{Yankowitz2019,Moon2014}. In addition, the wavelength of the moiré structure and the underlying potential modulation is well suited to observe the Hofstadter butterfly, a self-similar pattern of the energy levels as a function of a magnetic field \cite{ponomarenko2013,hunt2013}. 
\par

In contrast to hBN-SG systems, %
literature on hBN-BG and hBN-BG-hBN moiré structures is less extensive. In Ref.~\cite{Moon2014} the band structure of hBN-BG was calculated in a tight binding model that explicitly took the boron and nitrogen atoms into account. Then an effective continuum model was used to calculate the spectrum featuring Hofstadter’s butterfly. In Ref.~\cite{Zhang2019} $n$-layer structures built of graphene and hBN were analyzed theoretically, including interaction effects and a vertically applied electrical field, which contributes to the formation of flat bands and changes the Chern number. Further, correlated states such as quantum anomalous Hall insulating states and fractional anomalous quantum hall effect states were described.
In the experimental work of Ref.~\cite{Kim2018} band gap widths were measured at the central Dirac point and the secondary Dirac points both in hBN-SG and hBN-BG samples. It was found that in hBN-BG the secondary Dirac points are not fully gapped, and the particle hole asymmetry is less pronounced compared to  hBN-SG. The measurements were accompanied by theoretical calculations and it was concluded that the observed behavior can be explained qualitatively in a non-interacting picture.
Another experimental study created doubly aligned hBN-BG-hBN with close-to-zero rotation angle \cite{Endo2019}. They measured a non-local resistance which they attribute to a ``valley current'' near the band gap at charge neutrality.
In Ref.~\cite{Kuiri2021} doubly aligned hBN-BG-hBN was produced with a small twist in one hBN-layer and resistivity measurements displayed secondary Dirac points from both moiré structures. 
\par

In this work we want to understand hBN-BG and hBN-BG-hBN moiré structures more comprehensively by analyzing interaction-driven magnetic.
This magnetic order is relevant as a proxy for interaction-induced gaps in BG and %
is a promising 
candidate for the cases where such interaction-induced gaps may have been observed. 
In fact, in a number of experiments on freely suspended fewlayers \cite{Velasco,Bao,Freitag,Veligura,Bao} and multilayers \cite{Grushina,Morpurgo,Myhro} of graphene, all in Bernal stacking, evidence for an interaction-induced gap opening was found. This was interpreted theoretically as due to layered-antiferromagnetic order \cite{ZhangJung,Throckmorton,Kharitonov,Nilsson,Lang2012,SchererBi,SchererTri}. %
In theory, also SG hosts this ordering tendency \cite{honerkamp2008}, but the density of states at low energies is not as high as in the layered systems, and hence the ordering instability does not occur under normal circumstances. Clearly, it is of interest how this ordering tendency of layered graphene is altered in presence of an hBN environment. 
The influence of the aligned hBN is twofold%
: first, it modifies the band structure, which is addressed in this work.  Second, the hBN environment leads to additional screening of the electron-electron interactions. This effect is e.g.~%
dealt with by Roesner et al.~\cite{roesner2015} and can be incorporated in our reasoning by choosing an appropriately renormalized interaction constant.
\par
Our calculations for the interaction effects are restricted to charge neutrality and zero external magnetic field. %
First we extend the hBN-SG model of Ref.~\cite{Sachs2011} to a model of hBN-BG and hBN-BG-hBN. In this process we also distinguish between different stacking configurations. Next, we calculate the parameters for which the interacting spin susceptibility diverges in the random phase approximation (RPA). This provides information about the dominant magnetic order of these systems. Finally, we compare our RPA results to mean field calculations, where we are able to calculate the band gap widths of the magnetic phases. 

\section{Methods}
We set up a Hubbard model that describes the $\pi$-electrons of the carbon atoms, where each atom provides one orbital:
\begin{align}
H=&\sum\limits_{\substack{\vec{R}_1,\vec{R}_2,\\\vec{r}_1,\vec{r}_2,\sigma}}-t(\vec{d})c_{\vec{R}_1,\vec{r}_1\sigma}^\dagger c_{\vec{R}_2,\vec{r}_2\sigma} + 
U\sum\limits_{\vec{R}, \vec{r}}n_{\vec{R},\vec{r}\uparrow}n_{\vec{R},\vec{r}\downarrow}\nonumber\\
&+\sum_{\vec{r},\vec{R}}\epsilon^{\text{hBN}}_{\vec{r}}c_{\vec{R},\vec{r}\sigma}^\dagger c_{\vec{R},\vec{r}\sigma}
\end{align} 
The field operators $c_{\vec{R},\vec{r}\sigma}^{(\dagger)}$ annihilate (create) an electron with spin $\sigma$ at a site with lattice vector $\vec{R}$ and basis vector $\vec{r}$.
The atomic positions and the tight binding part including the hopping parameter $t(\vec{d})$, which is a function of the distance $\vec{d}$, is constructed as explained in the Refs.~\cite{DeTramblyLaissardiere2010, Moon2012}, a model originally used to describe %
twisted bilayer graphene. We fix the interlayer distance to be $d_{AB}=0.335\unit{nm}$ \cite{McCann2013}. The interaction is a Hubbard on-site interaction of strength $U$ and the operators $n_{\vec{R},\vec{r}\sigma}$ are the spin-dependent electron density operators. %
Although screening effects are less pronounced in graphene materials, and therefore nearest neighbor interaction is comparable to onsite interaction \cite{Rozhkov2016}, a description using on-site interaction only is justified when $U$ is properly renormalized \cite{schueler2013,Tang2015}.
\par
\begin{figure}%
	\centering
		\includegraphics[width=\linewidth]{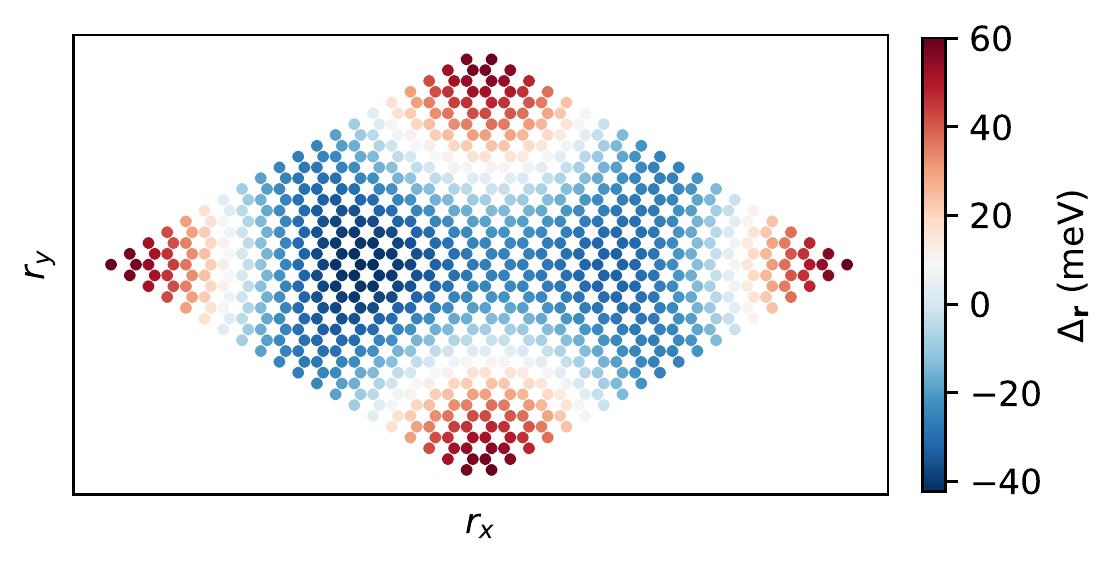}
		\caption{The `hBN-potential' $\Delta_{\vec{r}}$ acting on the carbon sites at position $\vec{r}=(r_x,r_y)$, as adapted from Ref.~\cite{Sachs2011}. The carbon sites are shown within one moiré unit cell with $N_{\text{moiré}}=20$.}%
		\label{pic84}
\end{figure}
\begin{figure}%
	\centering
	\includegraphics[width=\linewidth]{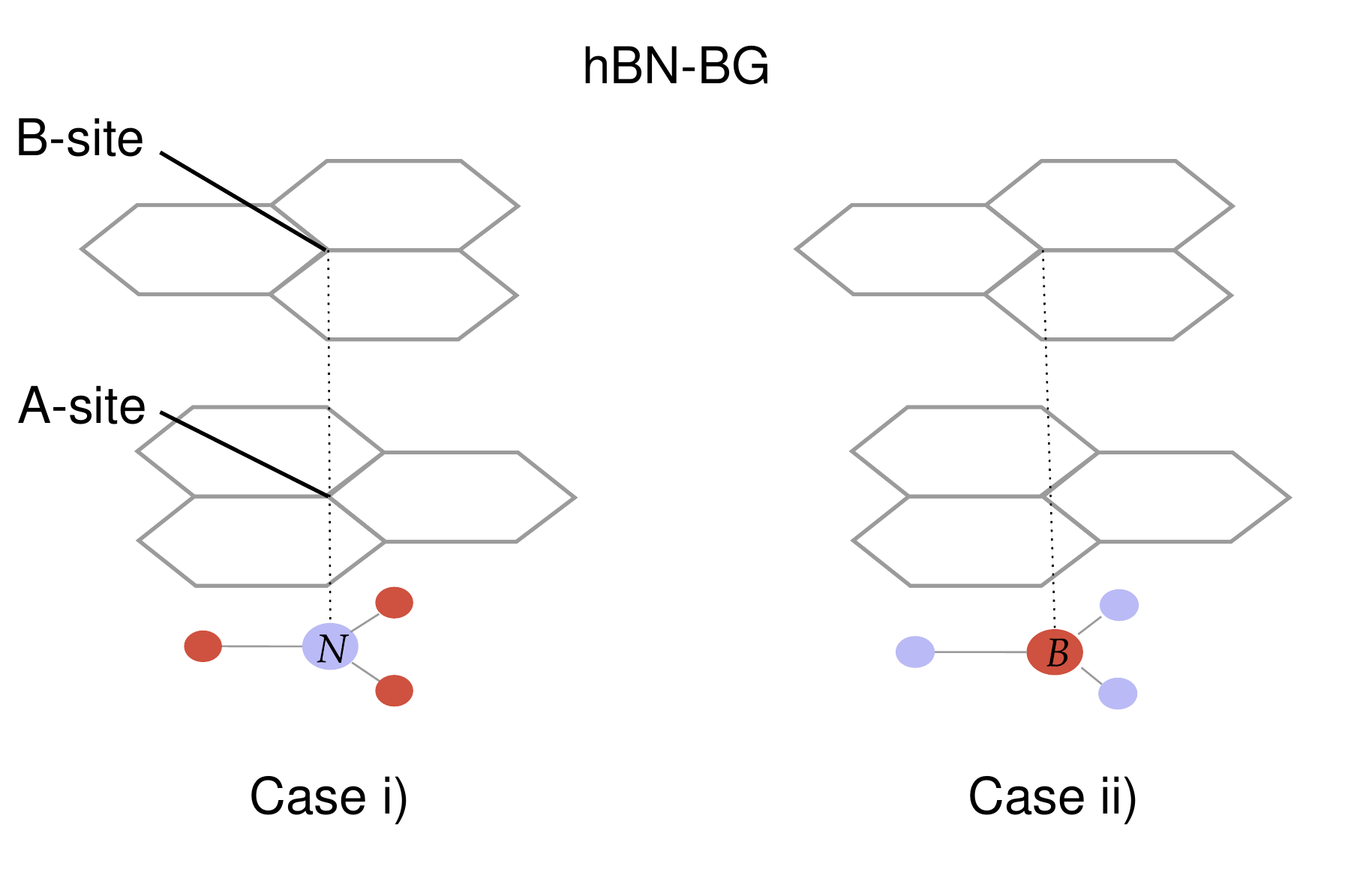}
	\caption{Different stacking configurations of hBN-BG. The second layer can be positioned such that at the corners of the moiré unit cell (see Fig.~\ref{pic84}) the stacking axis either goes through a nitrogen atom (case i) or a boron atom (case ii).}
	\label{tikz23}
\end{figure}
\begin{figure*}%
	\centering
	\includegraphics[width=\linewidth]{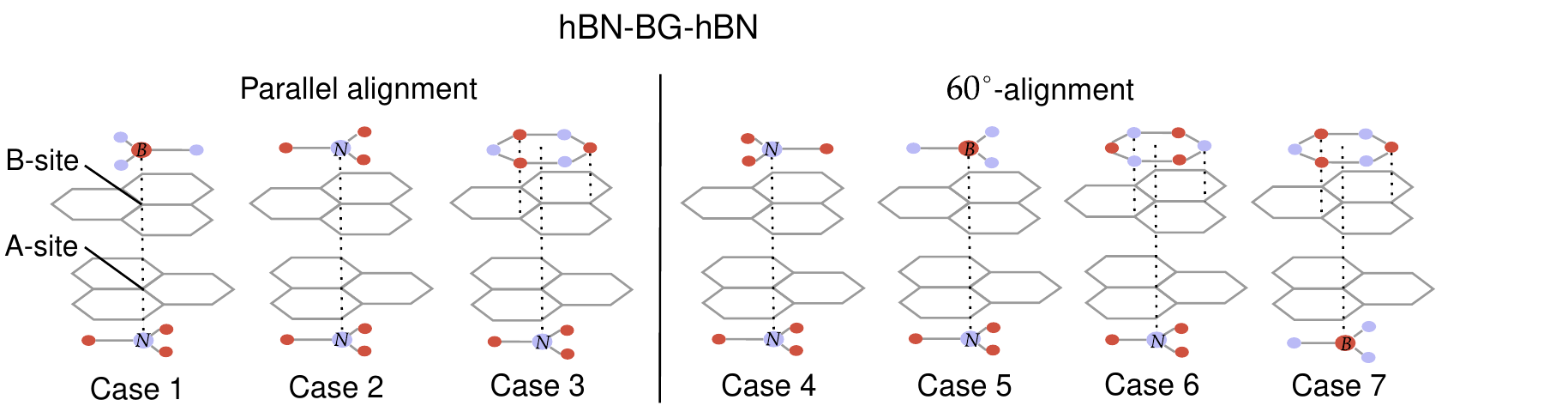}
	\caption{Different stacking configurations of hBN-BG-hBN. In the cases 1--3 the upper hBN layer is parallel to the lower hBN layer, in the cases 4--7 the upper hBN layer is rotated by $60^{\circ}$.}
	\label{tikz24}
\end{figure*}

The boron and nitrogen electrons are not taken into account explicitly in our model. Instead, the moiré structure is modeled by an effective potential term $\epsilon^{\text{hBN}}_{\vec{r}}$ on the carbon atoms, as proposed for hBN-SG in Ref.~\cite{Sachs2011}. There, a commensurate moiré structure is assumed with adjusted lattice constants to form a well-defined moiré unit cell.
A close approximation would be $\frac{a}{a_{\text{hBN}}}\approx \frac{55}{56}$ \cite{liu2003,Moon2014}. %
But since this requires a very large unit cell (with a lattice constant a factor $N_{\text{moiré}}=56$ larger than in graphene) and extensive computational resources, we perform most calculations on a simplified model with $\frac{a}{a_{\text{hBN}}}\approx \frac{19}{20}$ ($N_{\text{moiré}}=20$).
This means each unit cell in the moiré structure contains $N_{\text{moiré}}\times N_{\text{moiré}}\times 4$ carbon atoms. 
In Ref.~\cite{Sachs2011} an effective potential on the carbon atoms from the hBN-layer was proposed by performing a fit to DFT results. We repeated the last part of the fit using the same fit data points but a different function that keeps the threefold rotational symmetry: 
\begin{align}\label{eq:hBN_function}
\Delta_{\vec {r}}=A\,\Big(&\sin(\vec{G}_1\cdot\vec{r}+\phi) + \sin(\vec{G}_2\cdot\vec{r}+\phi)\nonumber\\
&+ \sin(-(\vec{G}_1+\vec{G}_2)\cdot\vec{r}+\phi)\Big) + C
\end{align}
Here, $\vec{G}_1$, $\vec{G}_2$ are the reciprocal primitive lattice vectors, and $A$, $\phi$ and $C$ are fit parameters, with best fit results at $A=P_A=21.9\unit{meV}$, $\phi=P_{\phi}=1.655$, and $C=P_C=-4.7\unit{meV}$. In the following, $\Delta_{\vec {r}}$ will be called hBN-potential and is shown in Fig.~\ref{pic84}. The effective potential incorporates $\Delta_{\vec{r}}$ with alternating sites on the two carbon atoms of the honeycomb lattice,
\begin{align}
&\epsilon_{\vec{r}}^{\text{hBN}}=s_{\vec{r}}\Delta_{\text{r}},
\end{align}
with $s_{\vec{r}}=+1/\shortminus1$  for  $\vec{r}$ being at an A-/B-site. %
To set up a model of hBN-BG from this model of hBN-SG,
we assume that the substrate only affects the first graphene layer \cite{Kim2018}, %
thus the hBN-potential is only added on one graphene layer. %
hBN-BG has two different stacking configurations, as pointed out in Refs.~\cite{Moon2014,Kim2018}. These two structures are shown in Fig.~\ref{tikz23}, which depicts one region of the moiré structure. %
For hBN-BG-hBN moiré structures we need to differentiate between more stacking configurations and we concentrate on those configurations where we can determine a clear stacking axis, as shown in Fig.~\ref{tikz24}. 
The hBN-potential is modified on the second graphene layer as to match a given configuration. For example, in hBN-BG-hBN case 3 the potential on the upper layer is multiplied by minus one and shifted by two third of the moiré cell diagonal.
\par
After having properly defined the non-interacting model, we can now take into account the Hubbard interaction and identify the possibility of magnetic order in the material. 
We calculate the magnetic susceptibility $\chi^m$ in RPA by summing up infinitely many diagrams via the exchange channel, analogous to Ref.~\cite{Klebl2019}, resulting in
\begin{align}
\hat \chi^m= \hat \chi \left({\mathbbm{1} +U\hat \chi}\right)^{-1}
\end{align}
Here, $\hat\chi$ is the polarization function\\
\begin{align}
&\chi_{\vec{r},\vec{r'}}(\omega,\vec{q})\nonumber\\
&=\frac{1}{\beta \Omega}\sum_{\omega'}\int \text{d}\vec{k}\ g_{\vec{r},\vec{r'}}(\omega',\vec{k}) g_{\vec{r'},\vec{r}}(\omega+\omega',\vec{k}+\vec{q}),
\end{align}
a matrix of the size of the number of atoms in the moiré cell. The functions $g$ are the bare green's functions, $\beta=\frac{1}{T}$ is the inverse temperature ($k_\text{B}=1$), $\omega$, $\omega'$ are bosonic and fermionic Matsubara frequencies, the integral is taken over wave vectors $\vec{k}$ in the Brillouin zone with area $\Omega$, and $\vec{q}$ is the wave vector of the corresponding magnetic spin density wave. 
The magnetic susceptibility diverges when the 'Stoner denominator', i.e. the matrix $\left({\mathbbm{1}+U\hat \chi}\right)$ to be inverted, develops a vanishing eigenvalue. This is the case for certain critical values of $T$ and $U$. Since the interaction strength is a model parameter that is not known exactly, we can calculate the critical interaction strength $U_c$ for fixed temperature via the relation $U_c=-\frac{1}{x_0}$. Here, $x_0$ is the most negative eigenvalue of the polarization function. $U_c$ marks the minimum interaction strength needed for the system to be in a magnetic phase at fixed temperature.
The eigenvector $\vec{x}_0$ corresponding to the eigenvalue $x_0$ is proportional to the magnetization on each atom, whose magnitude cannot be obtained from this analysis.
This is because all other eigenvectors become irrelevant when the spin susceptibility diverges.
As in Ref.~\cite{Klebl2019} we evaluate the  Matsubara sum in $\chi$ numerically and set $\omega=0$ for the static case and $\vec{q}=0$ for identical moiré cells. %
\par
We also calculate the magnetic order in mean field theory to compare our results with the RPA calculation. For this, we replace the spin density operator $S_{\vec{R},\vec{r}}^z=\frac{1}{2}(n_{\vec{R},\vec{r}\uparrow}-n_{\vec{R},\vec{r}\downarrow})$ by its mean field $S_{\vec{R},\vec{r}}^z\rightarrow \langle S_{\vec{r}}^z\rangle + \delta S_{\vec{R},\vec{r}}^z$ and neglect higher orders of fluctuations $\mathcal{O}((\delta S_{\vec{R},\vec{r}}^z)^2)$ \cite{Claveau2014}. %
Further, we neglect terms in the Hamiltonian describing charge density waves, because we want to focus on spin density fluctuations \cite{honerkamp2008}. The resulting non-interacting Hamiltonian (up to a constant) %
\begin{align}
H&=\sum_{\substack{\vec{r}_1,\vec{r}_2,\\\sigma,\vec{k}}}\Big(
H_{\vec{r}_1,\vec{r}_2}(\vec{k})- \sigma U \left\langle S_{\vec{r}}^z \right\rangle \delta_{\vec{r}_1\vec{r}_2} \Big)c^{\dagger}_{\vec{k},\vec{r}_1\sigma}c_{\vec{k},\vec{r}_2\sigma}
\end{align}
is then solved self-consistently. 
This allows us to obtain mean field estimates for interaction-induced energy gaps. These can then be compared with spectral gaps that derive from the 'single-particle'-description, here primarily from the substrate and stacking.  \\

\section{Results}
\begin{figure}%
	\centering
	\includegraphics[width=\linewidth]{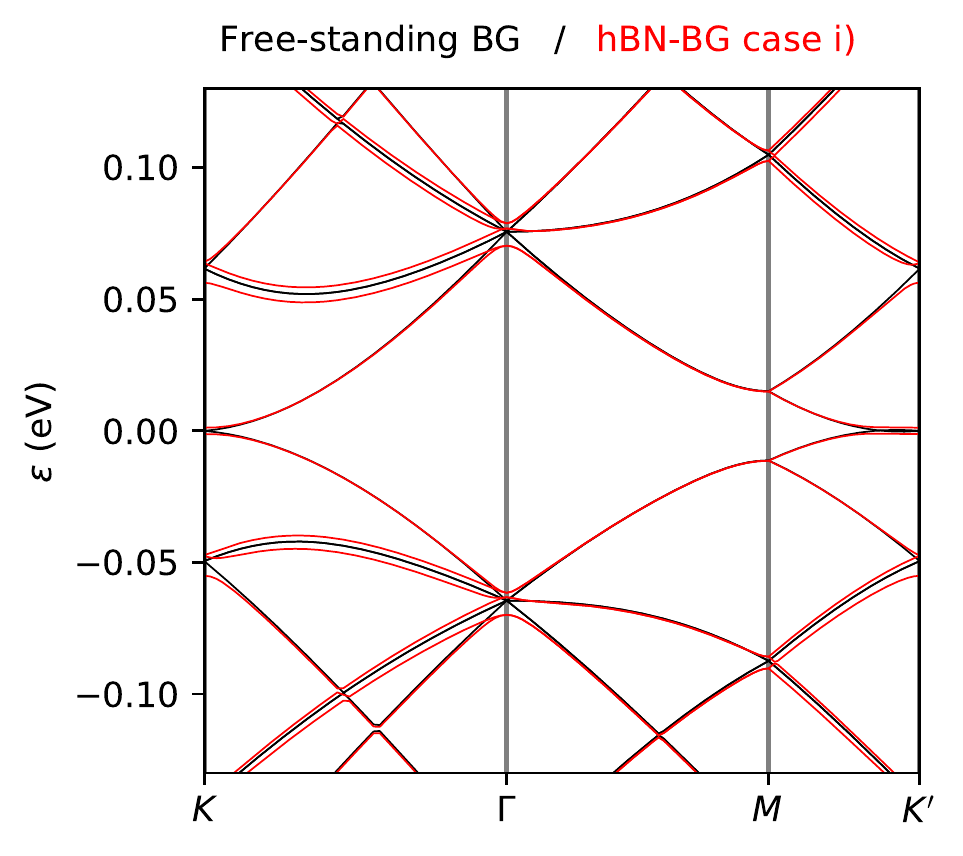}
	\caption{Band structure of hBN-BG ($N_{\text{moiré}}=56$, red lines) and free-standing BG (black lines) downfolded to the same mini Brillouin zone.  %
	At the $K$ and $K'$ points the hBN-potential (Eq.~\eqref{eq:hBN_function}) creates secondary, gapped Dirac points at band energies $\epsilon\sim+0.06\unit{eV}$ and $\epsilon\sim-0.05\unit{eV}$. In addition, a band gap opens at charge neutrality (see Fig.~\ref{pic86} for a zoom-in).}
	\label{pic85}
\end{figure}

Before we analyze the magnetic order, it is insightful to first understand the impact of the hBN-potential defined in Eq.~\eqref{eq:hBN_function} on the non-interacting system. In Fig.~\ref{pic85} we compare the band structure of hBN-BG, with the bands of (free-standing) BG downfolded into the mini Brillouin zone of the composite structure. 
In agreement with Ref.~\cite{Kim2018}, secondary Dirac points form at the new mini Brillouin zone edges and are split-up in hBN-BG. We observe only a small particle hole asymmetry. In addition, a band gap opens around zero energy.
\begin{figure}%
	\centering
	\includegraphics[width=\linewidth]{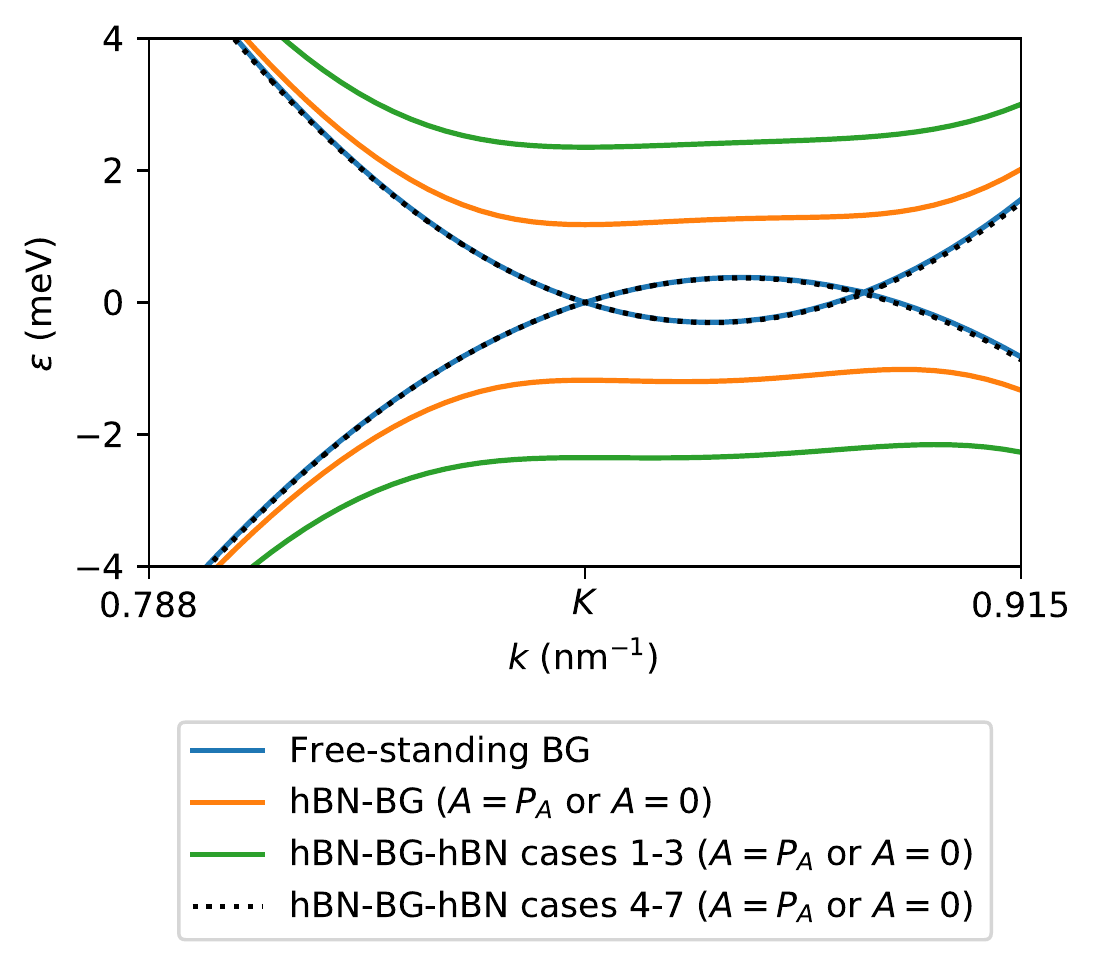}
	\caption{Low-enery band structure near the $K$-point of the mini Brillouin zone. The band structure shows a band gap at charge neutrality for hBN-BG and for hBN-BG-hBN cases 1--3. This band gap originates from an energy difference at neighboring carbon sites. Although alignment of the hBN layer and development of the moiré structure is necessary for the band gap opening, the gap itself can be modelled in a small unit cell corresponding to $A=0$ in the hBN-potential $\Delta_{\vec{r}}$.}
	\label{pic86}
\end{figure}
The low-energy band structure mostly determines our results of the magnetic order and is therefore shown as a zoom-in in Fig.~\ref{pic86}. In this figure we see two of the four Dirac points around $K$ in free-standing BG that are a result of trigonal warping \cite{Rozhkov2016}. 
In both configurations of hBN-BG a band gap opens. For the studied hBN-BG-hBN configurations the band structures split into two subgroups. In the cases 1--3 the band gap opens further compared to the one-sided cases whereas in the %
cases 4--7 the band gap closes again. The good agreement within the subgroups is a result of the fact that the $C$ parameter of the hBN-potential is mostly responsible for the band gap opening, while the $A$ parameter does not have a significant influence. The $C$ parameter causes an energy difference between A and B carbon sites in the cases 1--3 and an energy difference between dimer and non-dimer sites in the cases 4--7 \cite{muchakruczynski2010}. In Fig.~\ref{tikz24} we see that in the cases 1 
--3, the upper hBN layer is parallel to the lower layer (``parallel alignment''), whereas in the cases 4--7 the upper hBN layer is rotated by $60^{\circ}$ (``$60^{\circ}$-alignment'').%
\par

\begin{figure}%
	\centering
	\includegraphics[width=\linewidth]{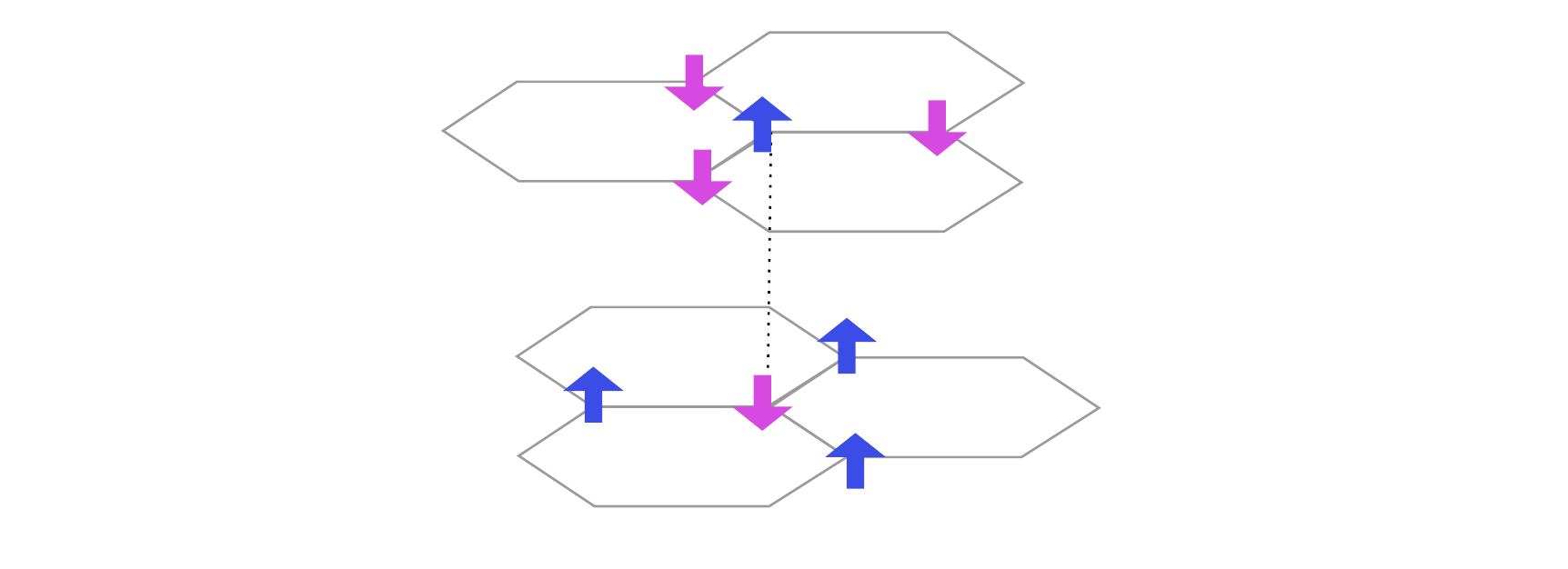}
	\caption{In all of our calculations the dominant magnetic order is found to be antiferromagnetic with opposite signs of the magnetization at neighboring carbon sites. Besides the relative spatial ordering pattern, the orientation of the magnetization is not fixed as long as the spin-orbit interaction is not taken into account. }
	\label{tikz25}
\end{figure}

Next, we turn to the possibility of magnetic order in our structures, %
within the RPA approach described above. In all cases studied, the dominant magnetic order is interlayer and intralayer antiferromagnetic, as shown in Fig.~\ref{tikz25}. 
No noticeable variation of this antiferromagnetic order was found within the moiré cell as the relevant eigenvectors turned out to be very homogeneous in their magnitudes through the moiré cell. This is quite different from the twisted bilayer cases studied, e.g.~in Ref.~\cite{Klebl2019,PhysRevB.102.085109}, where the eigenvectors vary strongly in the moiré cell.   
\begin{figure}%
	\centering
	\includegraphics[width=\linewidth]{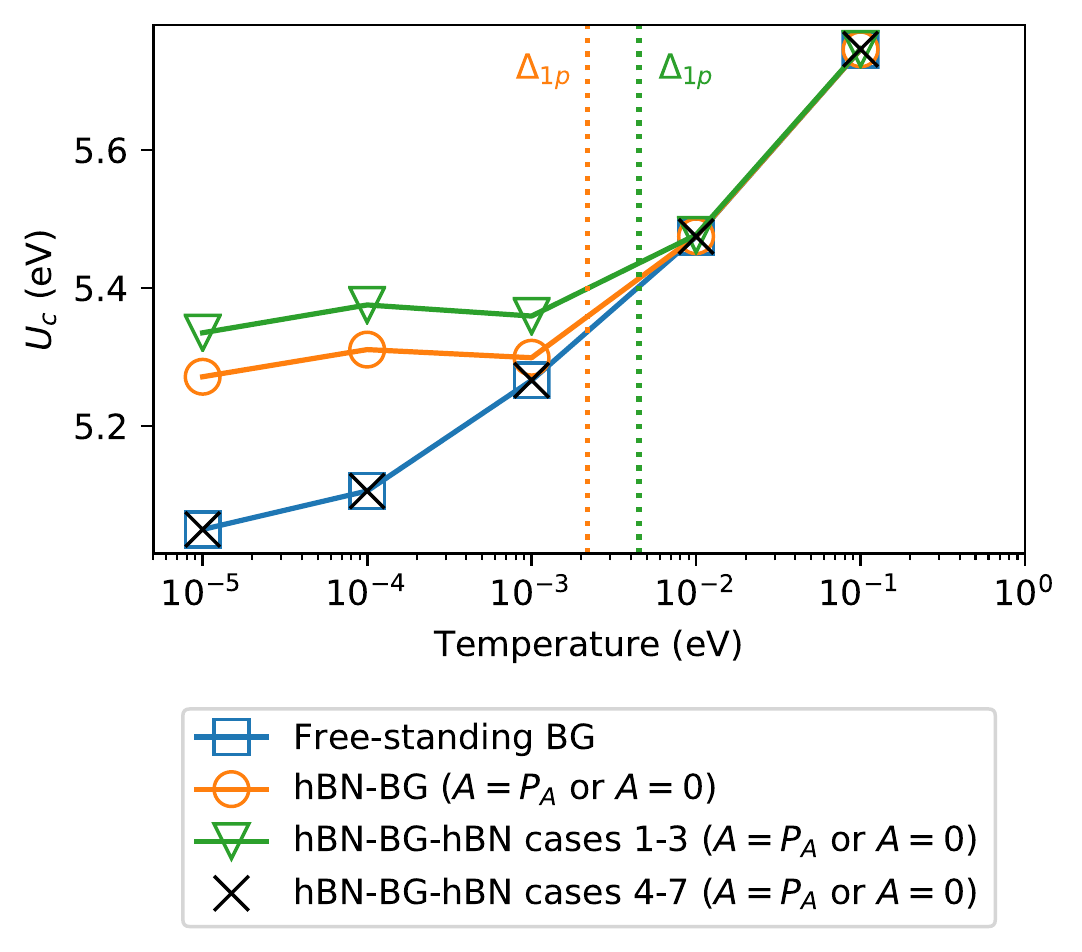}
	\caption{Critical Hubbard interaction $U_c$ for magnetic order as a function of the temperature. The different stacking configurations differ in that for structures with a band gap (e.g. hBN-BG) at temperatures smaller than the band gap, $U_c$ is larger as compared to structures without a band gap (e.g. free-standing BG).}%
	\label{pic87}
\end{figure}

Now, we calculate $U_c$ for the different structures as a function of the temperature in Fig.~\ref{pic87}. We find that the low-energy band structures of the non-interacting system fully determine the differences and the similarities between free-standing BG, hBN-BG and hBN-BG-hBN. 
For free-standing BG and $60^{\circ}$-aligned hBN-BG-hBN, structures with band crossings at the Fermi energy, $U_c$ continuously increases with the temperature. This is expected as thermal fluctuations at some point destroy the magnetic order. For `substrate-gapped' structures such as hBN-BG or parallelly aligned hBN-BG-hBN, $U_c$ differs from free-standing and $60^{\circ}$-aligned BG at temperatures below the size of the band gap. In this regime, $U_c$ saturates at an approximately constant value. This behavior is consistent with the expectations that band structure gaps become visible at temperatures below these energy scales.
Similar as in the analysis of the band structure, we see that the $A$ parameter of the hBN-potential does not play a role in our results. This means that, at least at charge neutrality, the moiré length scale does not influence the ordering tendency. 
In total, interaction effects are weakened by the substrate-induced band gap, although the difference in $U_C$ is small. 
\par
\begin{figure}%
	\includegraphics[width=\linewidth]{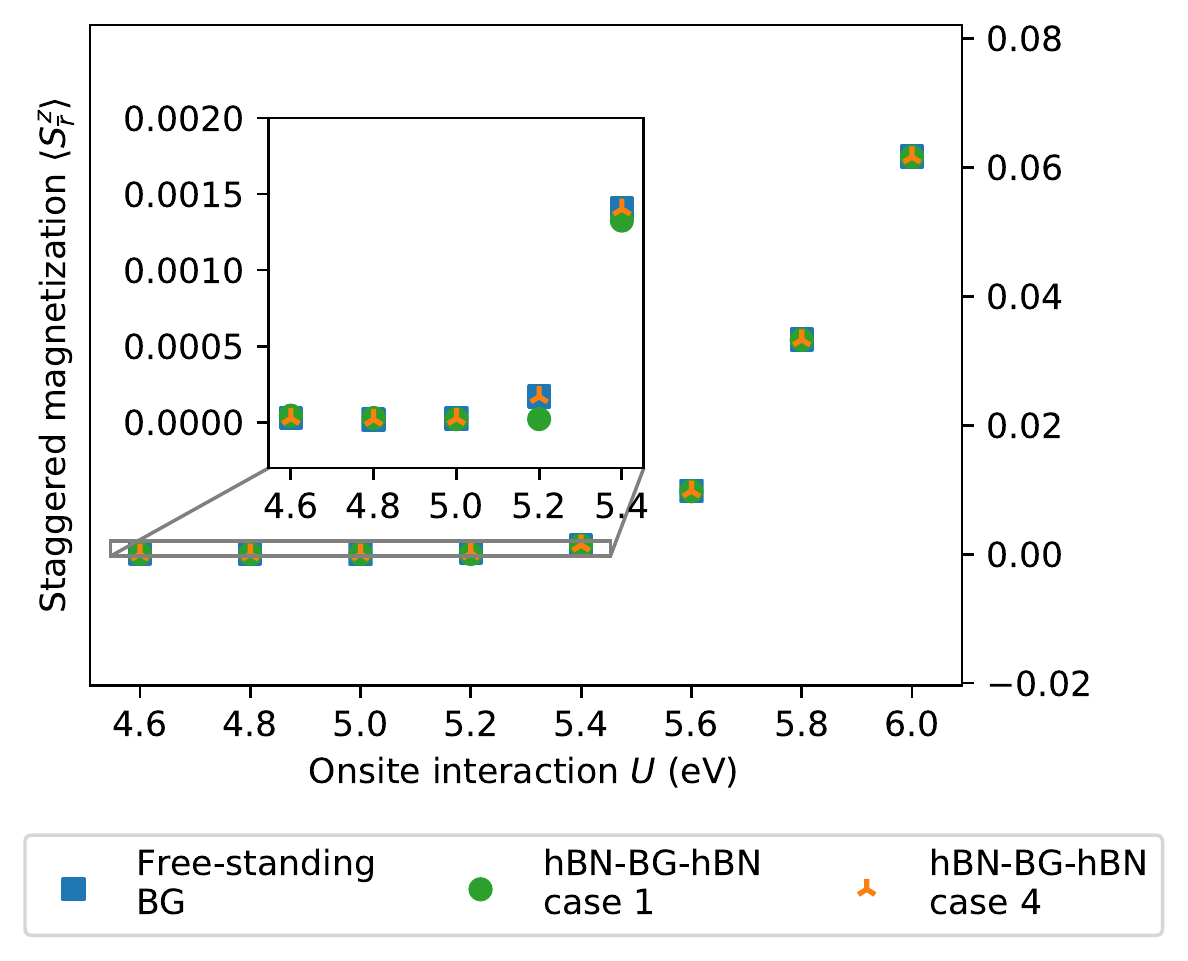}
	\caption{Staggered magnetization in self-consistent mean field theory as a function of the Hubbard interaction $U$ at a temperature of $T=10^{-4}\unit{eV}$. }
	\label{pic88}
\end{figure}
\begin{figure}
	\includegraphics[width=\linewidth]{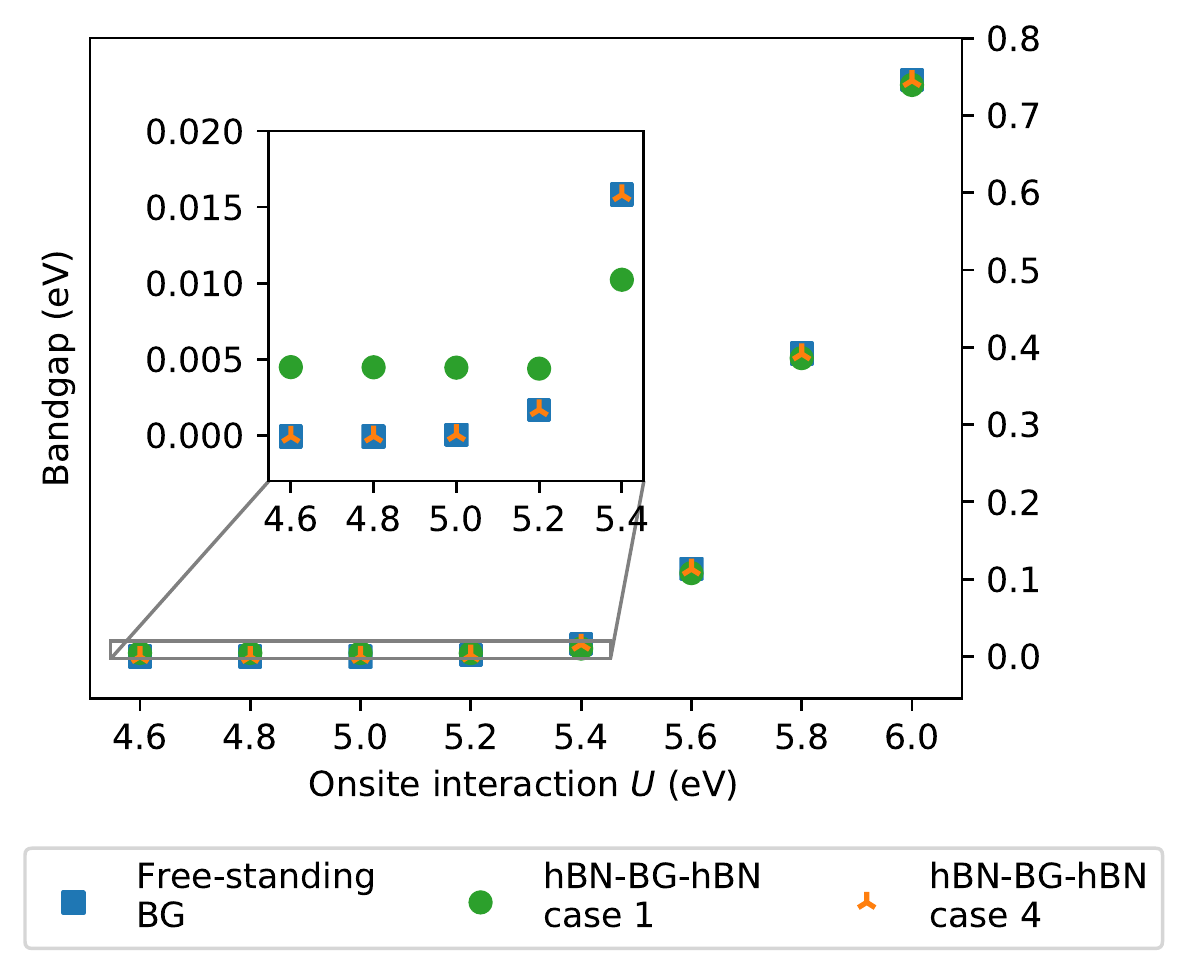}
	\caption{Band gap sizes of the converged mean field theory corresponding to Fig.~\ref{pic88} as a function of the Hubbard interaction $U$. For low $U$, the model can be described by the non-interacting system and only hBN-BG-hBN case 1 has a band gap. For large $U$ the magnetic state results in larger `magnetic' band gaps in all structures.}
	\label{pic89}
\end{figure}

Next, we evaluate the magnetic order of free-standing BG, hBN-BG-hBN case 1, and hBN-BG-hBN case 2 in mean field theory at a temperature of $T=10^{-4}\unit{eV}$. In Fig.~\ref{pic88} we show the magnetization $\langle S_{\vec{r}}^z\rangle$ in its magnitude and plot its absolute value averaged over the moiré cell as a function of $U$. The critical interaction strength $U_c$ needed to obtain a finite magnetization can be read off to be $5.0\unit{eV}<U_c<5.2\unit{eV}$ for free-standing BG and hBN-BG-hBN case 2, and $5.2\unit{eV}<U_c<5.4\unit{eV}$ for hBN-BG-hBN case 1. These values of $U_c$ are in agreement with the RPA results described above.
Next, we calculate the band structure of the converged mean field Hamiltonian. In Fig.~\ref{pic89} we extract the band gap widths at charge neutrality, again for different values of the interaction strength. As commonly observed in Hubbard models \cite{Claveau2014}, for parameters with nonzero magnetization an interaction-induced band gap is opened and we can see that the size of this  band gap increases with the magnetization. In the non-magnetic phase the structures are described by the non-interacting picture, which explains the band gap for low values of $U$ in hBN-BG-hBN case 1. At $U=5.2\unit{eV}$ this substrate-induced band gap in case 1 is still larger than the interaction-induced band gap in free-standing and hBN-BG-hBN case 2. 
At $U=5.4\unit{eV}$ the interaction-induced band gaps of free-standing BG and hBN-BG-hBN case 2 exceed that of hBN-BG-hBN case 1.


\section{Discussion}
Both, the RPA analysis and the mean field calculation show an upward shift of $U_c$ for substrate-gapped structures at low temperatures. 
Quantitatively, this shift is less than 6\%, putting constraints on the possibility that instabilities of BG may be observed in free-standing bilayers but not in the heterostructures. Even in this small window, the systems that do not have an interaction-driven ordered ground state show a substrate-induced gap. Thus, if no gap is observed, one should deal with a case without substrate-induced gap where the effective onsite repulsion is too small to trigger a ground state change at the given temperature. Lowering the temperature could still open a gap, but this gap might be too small to be observed. Note that the trigonal warping terms split any quadratic band crossing points of simple bilayer band structures into Dirac points, such that at zero temperature there should be a nonzero threshold value in $U$ for ordering.
\par

We also discuss the justification and magnitude of the onsite repulsion used in the Hubbard modelling of this work. Clearly, a more realistic model should, beyond the onsite term, also include non-local interaction terms.  This non-local interaction can approximately be incorporated in an effective onsite repulsion. In Ref.~\cite{schueler2013} such a simplification was analyzed in ab-initio calculations employing the constrained random phase approximation and it is suggested to estimate the effective onsite interaction via $U\approx U_0-U_1$, where $U_0$ is the `pure' onsite interaction and $U_1$ is the nearest-neighbor interaction. For free-standing BG a value of $U\approx9.3\unit{eV}-4.7\unit{eV}=4.6\unit{eV}$ would be obtained \cite{roesner2015}. However, the hBN-substrate may screen and thus reduce both, the `pure' onsite interaction and the nearest neighbor interaction as calculated in Ref.~\cite{roesner2015}. This screening from the substrate may approximately cancel out for the effective $U$ and we get, again, $U\approx 8.3\unit{eV}-3.7\unit{eV}=4.6\unit{eV}$. 
In all these approximations we stay well below our threshold values for ordering, i.e.~on the side where potential gaps would be induced by the hBN-layers. This is however not consistent with the experimental reports of gaps in freely suspended Bernal-stacked systems \cite{Velasco,Bao,Freitag,Veligura,Grushina,Morpurgo,Myhro}. Thus, we conclude that a quantitative description of these systems remains a challenge. 

\par
From the mean field calculation we find that, while the band gaps of single-particle nature are less variable, the interaction-induced gap is a strongly varying function of $U$ and largely independent of the putative non-interacting band structure once the interaction threshold is exceeded. This observation makes it difficult to compare our different materials, because the hBN-substrated materials may have a different effective $U$ compared to free-standing BG due to the additional screening effects.  Our mean field results imply that the precise value of $U$ will be the determining component for the size of the magnetic band gap.  Further, we see that hBN-BG and parellely aligned hBN-BG-hBN are gapped for all values of $U$. A case where the substrate destroys the ordering instability and the interaction-induced gap without producing another single-particle gap does not appear in our study.

\section{Summary}
In summary, we identified several stacking configurations and differentiated between parallelly aligned and $60^\circ$-aligned hBN-BG-hBN. The influence of hBN on the electronic structure of the BG was captured by an effective single-particle potential in analogy to earlier DFT work \cite{Sachs2011}. 
hBN-BG and parallelly aligned hBN-BG-hBN have a band gap of `single-particle' origin already in the non-interacting picture. For temperatures smaller than this band gap, the critical Hubbard interaction $U_c$ required for a magnetic ordering is larger than for the $60^\circ$-aligned hBN-BG-hBN cases and free-standing BG. Thus interaction effects are weakened by the substrate-induced band structure changes. However, the shift in $U_c$ is less than 6\%. 
Mean field theory allowed us to calculate the band gap sizes of the interacting systems. Our results demonstrate that the interaction-induced band gap is a strongly varying function of the onsite interaction $U$.

\subsubsection*{Acknowledgements}
The Deutsche Forschungsgemeinschaft (DFG, German Research Foundation) is acknowledged for support through RTG 1995, within the Priority Program SPP 2244 “2DMP” and under Germany’s Excellence Strategy-Cluster of Excellence Matter and Light for Quantum Computing (ML4Q) EXC2004/1 - 390534769. We acknowledge support from the Max Planck-New York City Center for Non-Equilibrium Quantum Phenomena. Calculations were performed with computing resources granted by RWTH Aachen University under project rwth0510.

\bibliography{literature}
\end{document}